\documentstyle[12pt]{article}
  \newcommand{\blankline}{\vskip .3cm}
  \newcommand{\f}{\begin{equation}}
  \newcommand{\ff}{\end{equation}}
\begin{document}
  \centerline{\Large  {\bf  General relativity with a topological phase:}}
  \centerline{\Large {\bf  an action principle}}
  \blankline
  \blankline
  \blankline
  \rm
  \centerline{{\bf Lee SMOLIN}$$,
  {\bf Artem STARODUBTSEV}$$}
  \blankline
    \centerline{\it $$Perimeter Institute for Theoretical Physics,
    Waterloo,  Canada}
\centerline{\it $$Department of Physics, University of Waterloo,
Waterloo,  Canada}
  \blankline
  \blankline
  \blankline
  \blankline
  \blankline
  \blankline
  \blankline

\centerline{ABSTRACT}

An action principle is described which unifies general relativity and topological
field theory. An additional degree of freedom is introduced, and depending on the value
it takes the theory has solutions that reduce it to 1) general relativity in Palatini form,
2) general relativity in the Ashtekar form, 3) $F\wedge F$ theory for $SO(5)$ and
4) $BF$ theory for $SO(5)$.  This theory then makes it possible to describe explicitly
the dynamics of phase transitions between
a topological phase and a gravitational phase where the theory has local degrees of freedom.
We also find that
a boundary between a dynamical and topological phase resembles an horizon.

\newpage
\tableofcontents

\section{Introduction}

One of the most important themes of recent work in non-perturbative approaches
to quantum gravity is the close connection between gravity and topological field theory.
One aspect of this is that many known gravitational theories can be expressed as {\it constrained
topological field theories}. These are theories in which the action is the sum of two terms-the first
described a topological field theory such as $BF$ theory and contains all the derivatives of the action,
while the second consists only of non-derivative constraints. The latter reduce the gauge invariance
of the topological field theory, as a result of which local degrees of freedom emerge. This
formulation is known to apply to general relativity in all dimensions, as well as $N=1$ and $N=2$
supergravity in $3+1$ dimensions and $d=11$ supergravity.

This connection makes possible a background independent and non-perturbative approach to
quantum gravity, at both the hamiltonian and path integral levels. The fact that the derivative
and boundary terms in the action of general relativity are shared by a $TQFT$, where exact
results can be achieved, means that the quantum theory of $GR$ or supergravity will
share the kinematical Hilbert space and path integral measure with
the $TQFT$, and differ from it only by the imposition of non-derivative constraints. As a result,
the study of spin foam models and loop quantum gravity has led to many exact results.

Other aspects of the gravity/$TQFT$ connection include the fact that there are physically
important classes of boundary conditions for $GR$, including those describing horizons, in
which the boundary term that must be added to the action is a Chern-Simons theory for
either $SU(2)$ or, in the case of non-vanishing cosmological constant, for the deSitter
or AdS group on the boundary\cite{}.  Yet another is the Kodama state, which appears to
be useful as a possible ground state of quantum gravity when the cosmological constant,
$\Lambda \neq 0$.

The connection with topological field theory naturally raises the question of whether there
might be dynamical transitions between a low energy phase in which gravity is approximately
described by general relativity, and a high energy phase, which is topological. Indeed,
speculations in this direction have been made for some time. For example, in 1988
Witten proposed that the Hagedorn temperature might represent a transition to a
topological phase in which the metric vanishes\cite{witten88}.
This kind of conjecture has recently
gained attention again in the context of spin foam models\cite{louis03}.

It has also been noticed that in certain first order action principles for
$GR$ and supergravity there are degenerate phases in which the determinate
of the metric vanishes\cite{degenerate}.
In \cite{ted97} phase boundaries were studied between regions in which
the metric is degenerate and non-degenerate, and were found to be null.

However, to describe a transition between a topological and a gravitational
phase dynamically, both must be solutions to the same theory.
In \cite{cdj} it was proposed that this could be done by making
the constraints that reduce the gauge symmetry of a $TQFT$ dynamical.
In this paper we would like to describe one way in which this can
be done.

In the next section we review the basic idea of gravity as
a constrained topological field theory and describe a new action principle
from which the constraints arise dynamically.
In section 3 we describe solutions to those constraints and show how
four different theories are recovered: Two $TQFT$'s: $F\wedge F$ theory
and $BF$ theory for $SO(5)$ and two versions of general relativity:
the action of Palatini and the action principle for the Ashtekar-Sen
variables. In section 4 we study the boundary between a topological
and gravitational phase and find that it resembles the conditions
imposed on an horizon.

\section{Action principle}
Our starting point will be a BF-theory action for SO(5) gauge
group. In this paper we consider a Euclidian theory, however
everything can directly be generalized for Lorentzian signature.
For writing down the action it is convenient to use
$\gamma$-matrices:
\begin{equation}
\gamma^A=\gamma^{A\dagger}, \ \ \{\gamma^A,\gamma^B\}
=\delta^{AB}, \label{gammadef}
\end{equation}
where $A,B=1,2...5$, and $\{.,.\}$ means anticommutator. Ten
generators of SO(5) group can then be represented as
\begin{equation}
J^{AB}=J^{AB\dagger}, \ \ \
J^{AB}=i[\gamma^A,\gamma^B].\label{jdef}
\end{equation}
15 matrices $\gamma^A$ and $J^{AB}$ form a basis in the space of
$4\times 4$ hermitian traceless matrices.

 The action principle for BF-theory
reads
\begin{equation}
S_{BF}=\int B^{AB}\wedge F^{CD} Tr \gamma_A \gamma_B \gamma_C
\gamma_D - \frac{\alpha}{2}\int B^{AB}\wedge B^{CD} Tr \gamma_A
\gamma_B \gamma_C \gamma_D. \label{bfa}
\end{equation}
Here
\begin{equation}
F^{AB}=dA^{AB}+A^A_C \wedge A^C_B
\end{equation}
is the curvature of SO(5)-connection $A^{AB}$ and $B^{AB}$ is an
arbitrary 2-form field. The equations of motion following from the
action (\ref{bfa})
\begin{eqnarray}
F^{AB}=\alpha B^{AB} \nonumber \\
\nabla \wedge B^{AB}=0
\end{eqnarray}
are trivially satisfied due to the Bianci identity $\nabla \wedge
F^{AB}=0$ and, therefore, the theory (\ref{bfa}) is topological.

By a small modification, however, which is a breaking of SO(5)
symmetry down to SO(4) the action (\ref{bfa}) can be turned into
that of General Relativity. General Relativity via symmetry
breaking was obtained in $F\wedge F$-theory by MacDowell and
Mansouri \cite{mmons} and in $BF$-theory by one of us
\cite{lee98}. For example one can insert a $\gamma_5$ in the trace
in the second term in (\ref{bfa})
\begin{equation}
S_{GR}=\int B^{AB}\wedge F^{CD} Tr \gamma_A \gamma_B \gamma_C
\gamma_D - \frac{\alpha}{2}\int B^{AB}\wedge B^{CD} Tr \gamma_A
\gamma_B \gamma_C \gamma_D \gamma_5, \label{gra}
\end{equation}
where $5$ labels some preferred direction. To see that the action
(\ref{gra}) is indeed the action of General Relativity let us
rewrite it in terms of $4+1$-decomposed indices $A=(\alpha,5)$,
$\alpha=1,2..4$:
\begin{equation}
S_{GR}=\int (B^{\alpha \beta}\wedge F_{\alpha \beta} + B^{\alpha
5}\wedge F_{\alpha 5}) - \frac{\alpha}{2}\int B^{\alpha
\beta}\wedge B^{\gamma \delta} \epsilon_{\alpha \beta \gamma
\delta 5}. \label{gra1}
\end{equation}
Also we can decompose the SO(5)-connection
\begin{equation}
A^{\alpha \beta}=a^{\alpha \beta}, \ \ A^{\alpha 5} =\frac{1}{l}
e^{\alpha}, \label{adecomp}
\end{equation}
where $a^{\alpha \beta}$ is an SO(4)-connection, $e^{\alpha}$ is a
tetrad, and $l$ is a constant of dimension of length.
(\ref{adecomp}) leads to the following decomposition of
SO(5)-curvature
\begin{eqnarray}
F_{\alpha \beta} &=& da_{\alpha \beta} +a_{\alpha \gamma} \wedge
a^\gamma_\beta+\frac{1}{l^2} e_{\alpha}\wedge e_\beta= f_{\alpha
\beta} + \frac{1}{l^2} e_{\alpha}\wedge e_\beta \nonumber \\
F_{\alpha 5}&=&\frac{1}{l} \nabla \wedge e_\alpha =T_\alpha.
\label{cdecomp}
\end{eqnarray}
Here $f_{\alpha \beta}$ is an SO(4)-curvature and $T_\alpha$ is a
torsion. The equations of motion for $B^{\alpha 5}$ from the
variation of (\ref{gra1})
\begin{equation}
F^{\alpha 5}=T_\alpha=0
\end{equation}
mean that the torsion is zero (which is the case for General
Relativity), and the equations of motion for $B^{\alpha \beta}$
\begin{equation}
F_{\alpha \beta} = f_{\alpha \beta} + \frac{1}{l^2}
e_{\alpha}\wedge e_\beta =\alpha B^{\alpha \beta}
\end{equation}
can be solved and substituted back into (\ref{gra1}) to yield
\begin{equation}
S_{GR}=\frac{1}{2G \Lambda} \int (f_{\alpha \beta} + \Lambda
e_{\alpha}\wedge e_\beta)\wedge(f_{\gamma \delta} + \Lambda
e_{\gamma}\wedge e_\delta) \epsilon^{\alpha \beta \gamma \delta},
\label{gra2}
\end{equation}
where $\Lambda=\frac{1}{l^2}$ is the cosmological constant and
$G={\alpha / \Lambda}$ is the Newton constant. The action
(\ref{gra2}) is the action of General Relativity.

It is possible to consider a theory in which the above symmetry
breaking is not introduced from the beginning but instead induced
by the theory itself. This is possible if e.g. the fixed quantity
$\gamma_5$ in (\ref{gra}) is replaced by a dynamical variable.

Let us consider the following action
\begin{equation}
S'_{GR}=\int B^{AB}\wedge F^{CD} Tr \gamma_A \gamma_B \gamma_C
\gamma_D - \frac{\alpha}{2}\int B^{AB}\wedge B^{CD} Tr \gamma_A
\gamma_B \gamma_C \gamma_D \Gamma +\int \lambda ( \Gamma^2-1),
\label{grma}
\end{equation}
where we introduced u(4)-valued (hermitian, but not necessarily
traceless) matrices $\lambda$ and $\Gamma$. For general covariance
$\Gamma$ should be a $0$-form (scalar) and $\lambda$ should be a
$4$-form (scalar density).

Let us first solve the equation for $\Gamma$ resulting from the
variation of the action (\ref{grma}) with respect to $\lambda$:
\begin{equation}
\Gamma^2=1. \label{eqgamma}
\end{equation}
As an hermitian $4 \times 4$ matrix $\Gamma$ can be represented as
\begin{equation}
\Gamma=u1+v_A \gamma^A + w_{AB}i[\gamma^A,\gamma^B],
\label{gammaexp}
\end{equation}
where $u$, $v_A$, and $w_{AB}$ are 16 arbitrary real numbers. By
substituting (\ref{gammaexp}) into (\ref{eqgamma}) and using the
anticommutation relations for $\gamma$-matrices (\ref{gammadef})
and
\begin{eqnarray}
\{i[\gamma^A,\gamma^B],\gamma^C\}&=&\epsilon^{ABCDE}i[\gamma_D,\gamma_E]\nonumber
\\
\{i[\gamma^A,\gamma^B],i[\gamma^C,\gamma^D]\}&=&\frac{1}{2}(\delta^{AC}\delta^{BD}-\delta^{AD}\delta^{BC})1
+\epsilon^{ABCDE}\gamma_E \label{gammaalg}
\end{eqnarray}
one finds
\begin{equation}
(u^2+v_Av^A+w_{AB}w^{AB})1+(u
v^E+\epsilon^{ABCDE}w_{AB}w_{CD})\gamma_E+(uw^{DE}+\epsilon^{ABCDE}w_{AB}v_C)i[\gamma_D,\gamma_E]=1
\end{equation}
This leads to the following set of equations for $u$, $v_A$, and
$w_{AB}$
\begin{eqnarray}
u^2+v_Av^A+w_{AB}w^{AB}&=&1 \nonumber \\
u v^E+\epsilon^{ABCDE}w_{AB}w_{CD}&=&0 \nonumber \\
uw^{DE}+\epsilon^{ABCDE}w_{AB}v_C&=&0 \label{uvweq}
\end{eqnarray}

\section{Solutions and phases}

In the absence of the general solution to the equations
(\ref{uvweq}) below we will give several examples. As
(\ref{uvweq}) is 16 non-linear equations for 16 parameters it is
natural to expect that different solutions to them are
disconnected from each other, i.e. cannot be transformed into each
other by a continuous change of parameters. The examples  are:

1. $u=1$, $v^A=0$, $w_{AB}=0$, which means that
\begin{equation}
\Gamma=1, \label{sol1}
\end{equation}
i.e. $4 \times 4$ unity matrix,

2. $u=0$, $w_{AB}=0$, and $v_A$ is an arbitrary 5-dimensional
vector such that $v_A v^A=1$, i.e.
\begin{equation}
\Gamma=\gamma^A v_A, \label{sol2}
\end{equation}

3.  $u=0$, $v^A=0$, and $w_{AB}$ is an antisymmetric tensor $5
\times 5$ such that all the nonzero components of it share one
common index and $w_{AB} w^{AB}=1$, which results in
\begin{equation}
\Gamma= i[\gamma^A,\gamma^B] w_{AB}, \label{sol3}
\end{equation}

4. $u=1/2$, $v^5=-1/2$, $w^{12}=w^{34}=1/2$, all the other
components being zero, i.e.
\begin{equation}
\Gamma=\frac{1}{2}(1-\gamma^5+i[\gamma^1,\gamma^2]+i[\gamma^3,\gamma^4])
\label{sol4}
\end{equation}

 Let us now see what kind of theories the above solutions result
in. If we plug the solution (\ref{sol1}) into the action
(\ref{grma}) then solve the equation for $B^{AB}$ and substitute
it back into the action we will obtain the $F\wedge F$ theory  for
SO(5) group.
\begin{equation}
S'_1=\frac{1}{2\alpha} \int F^{AB}\wedge F^{CD} Tr \gamma_A
\gamma_B \gamma_C \gamma_D. \label{res1}
\end{equation}
Due to the Bianci identity the bulk equations of motion of this
theory are trivial, and therefore the theory is topological.

If we use the solution (\ref{sol2}) in the action (\ref{grma}) we
will get the following result
\begin{equation}
S'_2=\int B^{AB}\wedge F_{AB}  - \frac{\alpha}{2}\int B^{AB}\wedge
B^{CD} \epsilon_{ABCDE}v^E. \label{res2}
\end{equation}
This action is very similar to the action (\ref{gra1}) except that
it includes an additional arbitrary parameter $v^A$. The
appearance of this parameter is an additional gauge freedom in the
action. This freedom can be fixed by aligning the vector $v^A$
along some preferred direction. Then the analysis
(\ref{gra1}-\ref{gra2}) can be repeated and the resulting action
will be the action of General Relativity. This is the ordinary
Palatini action for General Relativity which involves both
left-handed and right-handed connections.

We also have the solution (\ref{sol3}). After plugging it into
(\ref{grma}) the second term in the action will read
\begin{equation}
\frac{i\alpha}{2}\int B^{AC}\wedge B_C^B w_{AB}.\label{secondt}
\end{equation}
As $w_{AB}$ is antisymmetric and the tensor it is contracted with
is symmetric the contribution (\ref{secondt}) to the action
disappears. The resulting action
\begin{equation}
S'_{3}=\int B^{AB}\wedge F_{AB}   \label{res3}
\end{equation}
is the action of BF-theory for SO(5)-group. The equations of
motion of this theory mean that SO(5) curvature of the connection
$A^A_B$ is zero. So, although it is also a topological field
theory, it is slightly different from  the theory (\ref{res1}).

Finally, let us consider the solution (\ref{sol4}). The result
will be a sum of the results (\ref{res1}) and (\ref{res2}):
\begin{equation}
S'_4=\frac{1}{2\alpha} \int F^{AB}\wedge F_{AB}+\frac{1}{2\alpha}
\int F^{AB}\wedge F^{CD}\epsilon_{ABCD5}. \label{res4}
\end{equation}
This action is the self-dual part of the action of General
Relativity, which leads to the Ashtekar canonical formulation with
the Immirzi parameter equal to 1 (in the Euclidian theory
Ashtekar's variables are real). In the bulk the actions
(\ref{res2}) and (\ref{res4}) are equivalent as they differ from
each other by canonical transformation. However they may lead to
inequivalent field equations on the boundary. It is also
interesting to notice that this formalism doesn't seem to lead to
any other value of the Immirzi parameter than 1.

All these three solutions are disconnected from each other, so the
possible phase transition between them must be a first order phase
transition.

\section{Conditions on a phase boundary}
As it was mentioned the phase transition between different
solutions of the above theory is of the first order. Such
transition generally occur via formation of bubbles of a phase B
within a medium of a phase A. For two phases to coexist some
boundary conditions on a boundary between two phases must be
satisfied.

Consider a two-phase mixture of the above model one of which is
General Relativity (\ref{grma}) and the other is the Donaldson
theory (\ref{res1}). Let the phase boundary be located at $x_1=0$.
We dont specify whether the direction $x_1$ is spacelike or null.
With the traces of $\gamma$-functions calculated its action
principle reads
\begin{equation}
S_{2phase}=\frac{1}{2\alpha} \int F^{AB}\wedge F^{CD}
\Gamma_{ABCD}, \label{twoph}
\end{equation}
where
\begin{equation}
\Gamma_{ABCD}(x_1) =\delta_{AC}\delta_{BD} \theta(x_1)
+\epsilon_{ABCD5}\theta(-x_1), \label{bound}
\end{equation}
where $\theta{x_1}$ is the step $\theta$-function. The variation
of the action (\ref{twoph}) will give the equations of GR in the
region $x_1<0$, the equations of $F\wedge F$ theory (which are
trivially satisfied) in  $x_1>0$ region and the singular
contribution to the variation at $x_1=0$ resulting from
differentiation of $\theta$-functions
\begin{equation}
\delta S_{2phase}=...+\frac{1}{\alpha} \int F^{AB}\wedge n_1
\wedge \delta A^{CD} \delta(x_1)
\Delta_{ABCD}=...+\frac{1}{\alpha} \int_{x_1=0} F^{AB} \wedge
\delta A^{CD} \Delta_{ABCD}. \label{singt}
\end{equation}
Here $...$ stays for the bulk regular terms, $n_1$ is unit vector
in $x_1$-direction and
\begin{equation}
\Delta_{ABCD}=\delta_{AC}\delta_{BD}-\epsilon_{ABCD5}.
\end{equation}

For the variational principle to be well defined the singular term
(\ref{singt}) in the variation must vanish. For this the condition
$F^{AB}=0$ must be satisfied on a phase boundary, and, according
to the 4+1-decomposition (\ref{cdecomp}) this means

\begin{eqnarray}
f_{\alpha
\beta} + \Lambda e_{\alpha}\wedge e_\beta = 0\nonumber \\
T_\alpha=0. \label{bdecomp}
\end{eqnarray}
The second of the equations (\ref{bdecomp}) (zero torsion) is
always satisfied in GR, while the first equation is a specific
type of isolated horizons boundary conditions. This may suggest
that after a suitable generalization the dynamics of the formation
of a new phase will be governed by the dynamics of isolated
horizons.

As an example of further generalization of the theory given by
(\ref{grma}) one can consider the following action
\begin{equation}        
S'_{GR}=\int Tr B^{AB}\gamma_A \gamma_B\wedge (  F^{CD} T \gamma_C
\gamma_D+\beta d \Gamma \wedge d \Gamma) - \frac{\alpha}{2}\int
B^{AB}\wedge B^{CD} Tr \gamma_A \gamma_B \gamma_C \gamma_D
(\Gamma-\frac{1}{3}\Gamma^3) . \label{grma1}
\end{equation}
It is obvious that all the solutions of the theory (\ref{grma})
with constant $\Gamma$ considered above are also solutions of the
theory (\ref{grma1}).  (\ref{grma1}) may have other solutions, but
they would be difficult to analyze as that would require to
consider situations in which the equation of motion for $\Gamma$
is not 'decoupled' from those for other variables. In case of
solutions in which $\Gamma$ is varying in spacetime the term
containing derivatives of $\Gamma$ in  (\ref{grma1}) would define
the shape of the boundaries between domains of different phases.

\section*{Acknowledgments}
We thank Louis Crane, Laurent Freidel and Ted Jacobson for useful discussions.

\end{document}